\documentclass[a4paper,11pt]{article}
\pdfoutput=1
\usepackage{jheppub} 

\usepackage[font=small,labelfont=bf]{caption}
\usepackage[T1]{fontenc}
\usepackage[utf8]{inputenc}
\usepackage{epsfig}
\usepackage{graphicx}
\usepackage{xcolor}
\usepackage{latexsym}
\usepackage{textcomp}
\usepackage{amssymb}
\usepackage[colorlinks=true,linkcolor=blue,citecolor=blue, urlcolor=blue]{hyperref} 
\usepackage{amsfonts,amsthm,amstext,amscd}
\usepackage{amsmath}
\usepackage{mathtools}
\usepackage{upgreek}
\usepackage{bm}
\usepackage{comment}

\title{Testing the QCD formation time with reconstructed parton splittings}

\author[a]{Jasmine Brewer}
\author[b,c,d]{Wilke van der Schee}
\author[b]{Urs Achim Wiedemann}

\affiliation[a]{Rudolf Peierls Centre for Theoretical Physics, University of Oxford,  Oxford OX1 3PU, United Kingdom}
\affiliation[b]{Theoretical Physics Department, CERN, CH-1211 Gen\`eve 23, Switzerland}
\affiliation[c]{Institute for Theoretical Physics, Utrecht University, 3584 CC Utrecht, The Netherlands}
\affiliation[d]{Nikhef, Science Park 105, 1098 XG Amsterdam, The Netherlands}

\emailAdd{jasmine.brewer@physics.ox.ac.uk}
\emailAdd{wilke.van.der.schee@cern.ch}
\emailAdd{urs.wiedemann@cern.ch}

\abstract{
In high-energy elementary collisions the space-time ordering of parton branching processes is not accessible experimentally. In contrast, in heavy-ion collisions, parton showers interact with a spatially extended dense medium. This sets a reference length scale with respect to which the space-time ordering may be analysed. Here, we explore the possibility of identifying experimental signatures of the QCD formation time, $\tau_f$, on the level of a single parton splitting. Since heavy flavour offers an additional handle on tracing the propagation of individual quarks through the medium, we focus on the $g\to c\bar{c}$ splitting. Combining adapted versions of the Cambridge-Aachen and FlavourCone jet finding algorithms with grooming techniques, we show how the kinematics of such splittings can be reconstructed with high fidelity using either final state partons or hadrons, and how the formation time distribution of parton splittings can be constructed therefrom. Medium modification leads to a characteristic modification of this $\tau_f$ distribution. This effect can be used to construct experimentally-accessible ratios of $\tau_f$ distributions, in which the sensitivity of the medium modification to the QCD formation time becomes measurable.
}

\begin{document}
\preprint{CERN-TH-2025-053}

\maketitle

\section{Introduction}

In quantum field theory, all processes extend over finite regions of space and time. For radiative processes and for branchings in QCD parton showers, the formation time characterises this extension. In general, the formation time measures the duration of the process within which two daughter partons in a $1\to 2$ splitting decohere from each other and become independent quanta~\cite{Ioffe84,Dokshitzer:1991wu}.  However, in almost all QCD particle production processes studied in high energy physics, this formation time is not experimentally accessible. This is so since particle production in elementary interactions can be formulated entirely in momentum space without any reference to length or time scales. A parton shower in vacuum is simply a parton shower embedded in an environment that lacks a reference scale against which the formation time could be measured. 

The situation is different if instead the parton shower is embedded in the quark-gluon plasma (QGP) produced in ultra-relativistic nucleus-nucleus collisions. The lifetime and transverse size of the QGP  provides a suitable reference scale against which formation time can be measured. With the development of a partonic formulation of jet-QGP interactions, the formation time had been realized early on as an ordering principle for the space-time embedding of parton showers in the QGP,
see \cite{Eskola:1993cz,Zapp:2008af,Djordjevic:2009cr,Caron-Huot:2010qjx,Zapp:2011ya,Majumder:2013re,Modarresi-Yazdi:2024vfh} and Refs.~\cite{Cao:2020wlm, Cao:2024pxc} for recent reviews. This manuscript explores the opportunity to identify experimentally accessible signatures of the QCD formation time by reconstructing the kinematics of partonic $1\to 2$ splittings in the QGP.

\subsection*{The formation time of $1\to 2$ splittings in a medium}
 To make these considerations quantitative, it is instructive to consider a simplified setting in which a partonic $1\to 2$ splitting interacts with the QGP via a single gluon exchange. Varying the distance $\xi$ between the position of the scattering center in the QGP 
 and the position at which the parent parton is produced, one can then discuss how the medium modification of the splitting depends on its spatio-temperal embedding in the QGP. 

Calculations in this simplified set-up have been performed for the medium modifications of all $q\to qg$, $g\to 
gg$~\cite{Wiedemann:2000za,Gyulassy:2000er,Salgado:2003gb} and  $g\to q\bar{q}$~\cite{Kang:2016ofv,Attems:2022ubu} splitting functions in the so-called $N=1$ opacity expansion of a single scattering. For the case that the position $\xi$ of the scattering center is homogeneously distributed within an in-medium path length $L$ with a uniform density $n(\xi) = n_0\, \Theta(L-\xi)$, the medium modification of the splittings take the simple form
\begin{eqnarray}
\left(P_{1 \to 2}\right)^{\rm med}_{N=1}
	&\propto&  n_0\, L \int \frac{d{\bf q}}{(2\pi)^2} \sigma_{\rm el}({\bf q},z) \, 
	\nonumber \\
	&& \times \left(  1 - \frac{\tau_f}{L} \sin\left[ L / \tau_f\right]   \right)
\left[\hbox{\rm integrand}   \right]\,.
		   \label{eq1}
\end{eqnarray}
Here, $\sigma_{\rm el}({\bf q},z)$ is the differential elastic cross section
for the $1\to 2$ splitting to exchange transverse momentum ${\bf q}$ with a constituent in the medium. This cross section can depend on the longitudinal momentum fractions $z$ and $(1-z)$ of the splittees. The opacity $n_0\, L\, \sigma_{\rm el}$ in the first line of Eq.~\eqref{eq1} determines the overall probability that the $1\to 2$ splitting interacts with the QGP. The $\left[\rm integrand \right]$ is a somewhat involved combination of kinematic factors that differs for different $1\to 2$ splittings, but does not depend on the in-medium path length $L$. The focus of the present work will be on the first factor in the second line of Eq.~\eqref{eq1}.  This factor makes explicit that for fixed total interaction probability (i.e. for fixed $n_0\, L \sigma_{\rm el}$), what happens depends on both the momentum transfer ${\bf q}$ between the QGP and the $1 \to 2$ splitting, and the distance between the position at which the parent parton is produced and the position at which the momentum is transferred. This non-trivial distance dependence is encoded in a factor which interpolates between totally incoherent and totally coherent limiting cases
\begin{equation}
    \left(  1 - \frac{\tau_f}{L} \sin\left[ L / \tau_f\right]   \right) \longrightarrow
    \begin{cases} 1 &\hbox{\rm for}\quad L \gg \tau_f \\ 0 &\hbox{\rm for}\quad \tau_f \gg L\end{cases}\, .
    \label{eq2}
\end{equation}
For all medium-modified splitting functions in the $N=1$ opacity expansion, $\tau_f$ in the previous expression takes the form
\begin{equation}
    \tau_f = \frac{2E}{Q_1^2}\, ,
    \label{eq:tauform}
\end{equation}
where $E$ is the energy of the parent parton in the QGP rest frame and $Q_1^2$ is the squared invariant mass (virtuality) of that parent parton {\it prior} to interacting with the medium. We note that already the earliest discussions of the space-time picture of QCD bremsstrahlung in the vacuum~\cite{Ioffe84,Dokshitzer:1991wu} arrived at formation times of the form  \eqref{eq:tauform}.

Though $\tau_f$ in Eq.~\eqref{eq:tauform} is written in terms of momentum space variables, several lines of argument suggest that it can be associated with the spatial and temporal location where the splitting was formed, namely its formation time. First, Eqs.~\eqref{eq1} and \eqref{eq2} make it clear that, with the overall interaction probability $n_0\, L \sigma_{\rm el}$ fixed, the medium modification of a $1 \to 2$ splitting depends on comparing the $\tau_f$ of that splitting to the physical length scale $L$ of the medium\footnote{All results discussed here have been derived in a close-to-eikonal approximation, and we therefore identify longitudinal spatial distance and time in our arguments.}. Second, parametrically, Eq. \eqref{eq:tauform} is consistent with the intuitive estimate that a parent parton of initial virtuality $Q_1$ decays on a time-scale $\tfrac{1}{Q_1}$ in its rest frame. If that timescale is measured in the local rest frame of the QGP, it is time-dilated by a Lorentz factor $\tfrac{E}{Q_1}$ which yields $\tfrac{E}{Q^2_1}$~\cite{Dokshitzer:1991wu}. Third, if one inspects in detail how the factor $\tau_f$ arises in the derivation of medium-modified splitting functions~\cite{Wiedemann:2000za,Gyulassy:2000er,Kang:2016ofv,Attems:2022ubu}, one finds that configuration space information enters these calculations through phases $\exp\left[ i E_i t  - i p_i^L x_L\right]$ associated to the parton $i$. In the eikonal approximation when $t = x_L$ and $p_i^L = \sqrt{E_i^2 + {\bf k_i}^2} \approx E_i - \tfrac{{\bf k_i}^2}{2E_i}$, these phases are given in terms of transverse energies times longitudinal positions $x_L$, namely
$\exp\left[ i E_i t  - i p_i^L x_L\right] \approx \exp\left[ i \tfrac{{\bf k_i}^2}{2E_i} x_L\right]$. The formation time \eqref{eq:tauform} arises from sums of such transverse energies $\tfrac{{\bf k_i}^2}{2E_i}$.
In this sense, $1/\tau_f$ is the Fourier conjugate of a longitudinal distance in the medium, which justifies the configuration space interpretation of $\tau_f$. This third line of argument has two further noteworthy implications: On the one hand, it explains the factor $2$ in Eq.~\eqref{eq:tauform} that arises in explicit calculations but that cannot be obtained by parametric reasoning alone. On the other hand, configuration space information enters via phase factors of the same form $\exp\left[ i E_i t  - i p_i^L x_L\right]$
in all derivations of medium-modified splitting functions, and not only in derivations within the $N=1$ opacity expansion. This explains why even in refined formulations in which the medium modification is not of the form \eqref{eq1}, the formation time $\tau_f$ in Eq.~\eqref{eq:tauform} is the physically-relevant time scale for the splitting process.  

\subsection*{The utility of $g\to c\bar{c}$ for accessing the formation time}

The goal of the present work is to identify direct, experimentally-accessible tests of the formation time of splittings in the medium. 
There are several other recent works that have explored specific classes of observables with similar motivation. For instance, one may try to determine a proxy for a time axis by reclustering entire jets. This allows one to label jets according to their fragmentation pattern and to select jet populations with enhanced sensitivity to quenching effects~\cite{Apolinario:2024hsm}. Another idea is to make use of the hadronic decays of boosted W bosons in top-quark decays in future experiments to implant in the medium what is essentially an electro-weak initiated but time-delayed jet~\cite{Apolinario:2017sob}. Here, we search for experimentally accessible signatures of formation time on the level of individual parton splittings that could be arguably more direct and more abundant than previously proposed measures. 

Ideally, we would like to identify individual $1 \to 2$ splittings in the medium and quantify how their modification depends on $\tau_f$. This necessitates reconstructing the kinematics of splittings with sufficient fidelity from the hadronic final state to become sensitive not only to the medium modification of $1\to 2$ splittings, but also to the $\tau_f$-dependence of that medium modification. In this respect, the $g \to c\bar{c}$ splitting offers several key qualitative advantages compared to other splitting functions.

First, heavy flavour is conserved on QCD timescales, meaning that one can reliably associate charm hadrons in the final state with charm quarks produced in the parton shower. Over the last decade, jet substructure has proven to be a key tool that can provide access to the kinematics of $1 \to 2$ splittings~\cite{Larkoski:2017bvj}. However, it has a key limitation that it cannot identify the parton flavour of the splitting. To our knowledge, a gluon splitting to heavy quarks provides the only situation where one can use jet substructure to isolate an individual type of splitting, particularly beyond the first (groomed) splitting in the shower. Since splittings with sizeable formation times need not be the first splitting, this is a key feature for our study.

Second, the charm mass $m_c \sim 1.3\,$GeV$/c^2$ provides a particularly interesting scale. On the one hand, the overall size of the medium modification is $\propto \left(\tfrac{\hat{q}L}{m_c^2}\right)$, a scale that is order unity (if not larger) for typical values of transverse momentum broadening $\hat{q}L$ in the medium~\cite{Huss:2020whe,JET:2013cls}. As a consequence, the calculated medium modification of this splitting is expected to be sizeable enough to be in experimental reach~\cite{Attems:2022otp}. On the other hand, as we shall demonstrate in the following, $m_c^2$ is sufficiently small that one can collect samples of $g\to c\bar{c}$ in heavy ion collisions at the LHC whose formation times are comparable to the in-medium path length. For beauty quarks, both criteria are more problematic: the medium modification of the $g\to b\bar{b}$ splitting is a factor $m_c^2/m_b^2 \approx 1/10$ smaller than that of $g\to c \bar{c}$~\cite{Attems:2022otp}. The smallest invariant mass of the $b\bar{b}$-pair is also a factor  $m_b^2/m_c^2 \approx 10$ larger, which means that it is much more difficult to identify splittings with formation times that are sufficiently long to be comparable with the system size. 

Third, the absence of a soft singularity in both medium-modified and vacuum $g\to c \bar{c}$ splitting functions facilitates control over grooming cuts that ensure that both daughter prongs are sufficiently hard. As will be seen explicitly in the following, the kinematic reconstruction of both daughter prongs of a parton splitting is indispensable for a reliable determination of formation time. Such reconstruction would be more complicated if the prong is soft.

\subsection*{Medium modification of the $g \to c\bar{c}$ splitting}

For the reasons outlined above, we focus in the following on the $g \to c\bar{c}$ splitting function. We emphasise, however, that we are exploring a formation time dependence that is generic. The medium modification of {\it all} parton splittings is expected to depend on a QCD formation time of the form of Eq.~\eqref{eq:tauform}, and the same formation time-dependence has been found in calculations based on different phenomenological assumptions and different technical approximations. For instance, the same scale interpolates between totally coherent and incoherent medium modification of $g\to c{\bar c}$ if the medium interacts not via $N=1$ but via multiple soft scatterings (in the so-called harmonic oscillator approximation) ~\cite{Caron-Huot:2010qjx,Attems:2022ubu}, and comparable medium modifications are found in both approximation schemes. As the assumption of multiple soft momentum transfers is expected to be closer to the phenomenological reality, we shall base our numerical studies in Section~\ref{secmed} on this multiple-soft scattering formula. However, in the path-integral formulation of that multiple soft scattering approximation, the $\tau_f$-dependence reveals itself only after some non-trivial analytical manipulation or numerical calculation. For the sake of a transparent and explicit discussion, we are therefore using in this introduction the simple and explicit $N=1$ opacity expression to exemplify the role of $\tau_f$.  To first order in opacity, the medium modification of the $g \to c\bar{c}$ splitting function takes the form~\cite{Kang:2016ofv,Attems:2022ubu}, 
(see Refs.~\cite{Caron-Huot:2010qjx, Kang:2016ofv, Sievert:2019cwq, Attems:2022ubu} for results beyond the $N=1$ opacity expansion)
\begin{eqnarray}
	\left(\frac{1}{Q^2}\, P_{g \to c\, \bar{c}} \right)^{\rm med}_{N=1}
	&=& \frac{1}{2} n_0\, L \int \frac{d{\bf q}}{(2\pi)^2} \sigma_{\rm el}({\bf q},z) \, 
	\left(  1 - \frac{\tau_f}{L} \sin\left[ L / \tau_f\right]   \right)
	\nonumber \\
	&& \times \left[	 \left(\frac{1}{Q^2}\, P_{g \to c\, \bar{c}} \right)^{\rm vac}_{\bm{\upkappa} \to \bm{\upkappa} + {\bf q}}  -	 
  \left(\frac{1}{Q^2}\, P_{g \to c\, \bar{c}} \right)^{\rm vac} 
  \right.
	\nonumber \\
 && \quad  + \left.	\frac{(Q^2-Q_1^2)^2 m_c^2}{2 Q^4 Q_1^4 z(1-z)}
	+ \left( \frac{\left(\bm{\upkappa}
		   + {\bf q} \right)}{Q_1^2} - \frac{\bm{\upkappa}}{Q^2}  \right)^2  \frac{z^2 + (1-z)^2}{2z(1-z)} \right]\,.
		   \label{eq4}
\end{eqnarray}
This expression is clearly of the form of Eq.~\eqref{eq1}, and all comments made about formation time in the preceding section therefore apply. The integrand depends on the transverse momentum ${\bf q}$ transferred from the medium to the splitting and it depends on the transverse momentum 
\begin{equation}
  \bm{\upkappa}  = {\bf p}_c (1-z) - {\bf p}_{\bar{c}} z \, ,  \label{eq5}
 \end{equation}  
that is defined in terms of the observed final transverse momenta ${\bf p}_c$ and 
$ {\bf p}_{\bar{c}}$ of the charm and anti-charm quark, and their longitudinal momentum fractions $z$ and $(1-z)$, respectively. In the close-to-eikonal approximation in which Eq.~\eqref{eq4} was derived, the squared invariant masses $Q^2$ and $Q_1^2$ of the $c\bar{c}$ pair in the final state and prior to interaction with the medium take the explicit forms
\begin{equation}
    Q^2 = \frac{\kappa^2 + m_c^2}{z(1-z)}\, ,\qquad
    Q_1^2 = \frac{(\kappa+{\bf q})^2 + m_c^2}{z(1-z)}\, .
    \label{eq6}
\end{equation}
Physically, the second line of Eq.~\eqref{eq4} represents the medium-induced broadening $\bm{\upkappa} \to \bm{\upkappa} + {\bf q}$ of a  charm-anticharm pair that formed with relative momentum $\bm{\upkappa}$ in the vacuum. The last line is a manifestly-positive contribution that represents medium-induced $c\bar{c}$-radiation~\cite{Attems:2022otp} from gluons propagating through the QGP. 

In the simulations shown in later parts of this work, we will use the exact expressions for  $Q$ and $Q_1$ rather than the high-energy approximation~\eqref{eq6}. Also, we shall analyse the kinematic distribution of $c\bar{c}$ pairs in the transverse centre of mass frame of the pair, in which we denote ${\bf k}_T\equiv {\bf p}_c = -{\bf p}_{\bar{c}}$ and hence $\bm{\upkappa} =  {\bf k}_T$.

\subsection*{A roadmap to establishing experimental sensitivity to the formation time}

The aim of this manuscript is to establish an experimentally-testable relation between Eq.~\eqref{eq:tauform} measured in {\it momentum space} and the {\it configuration space} interpretation of this ratio in terms of a longitudinal distance or time $\tau_f$.
If one would measure only splittings that occur in the vacuum, one could construct Eq.~\eqref{eq:tauform}, of course, but the space-time interpretation of $\tau_f$ would not be testable due to the absence of an experimentally-controllable length scale.
In contrast, if the same vacuum splittings are embedded in a dense QCD medium of length $L$, then we aim to show that the medium-modification is restricted to the fraction of the event sample that satisfies $\tau_f < L$, and that this fraction can be varied with the scale $L$ that can be controlled experimentally. It is the experimental sensitivity to the interference term Eq.~\eqref{eq2} that establishes an experimental sensitivity to the formation time. The layout of this manuscript follows the resolution of three critical issues that need to be clarified to follow such a strategy:

\begin{enumerate}
\item {\it Reconstructing with high fidelity the partonic $1\to 2$ splitting from a hadronic final state.}\\
To determine $\tau_f$ for a particular splitting, it is necessary to reconstruct the energy and virtuality of the parent parton from the hadronic final state. More generally, we aim to reconstruct \textit{all} kinematic information of $g \to c\bar{c}$ splittings, i.e., the energy $E$ of the parent parton, the longitudinal momentum fractions $z$ and $(1-z)$ of the two daughters, and their relative transverse momenta. 
In Section~\ref{sec:reconstruction} we outline two strategies to perform this reconstruction and demonstrate their accuracy in Monte Carlo studies.
\item {\it Identifying kinematic ranges in which $\tau_f$ is of the same order as the QGP lifetime.}\\
For our aim, it is further necessary to demonstrate that it is possible to select samples of splittings whose formation times can be comparable to the in-medium path length. This is challenging for several reasons. For a gluon splitting that results in a pair of $D^0$ mesons, we have that $Q > 2m_D > 3.7\,$GeV$/c^2$. Therefore splittings below $50 \text{ GeV}$ have $\tau_f < 1.44\,$fm$/c$, which is much shorter than the typical QGP in-medium length of about $5\,$fm$/c$. In addition, many splittings occur at sizeable transverse momentum, resulting in even shorter typical formation times. In Section~\ref{sec3A} we demonstrate how splittings with low virtualities in jets of $50 - 300$~GeV can be used to scan the formation time over the relevant range of length scales of the QGP, $ 1 \, {\rm fm} < \tau_f <  5\, {\rm fm}$. We also discuss the limitations of this reconstruction, especially arising due to hadronisation effects in this regime.
\item {\it Establishing sensitivity to the $\tau_f$-dependence of medium-modified $1\to 2$ splittings. }\\ 
Finally, in Section~\ref{sec3b} we provide evidence in a model study that the change of the medium modification with $\tau_f$ within the experimentally-accessible $\tau_f$ range may be sufficiently large to be experimentally accessible. It is this third step that ultimately provides evidence for possible experimental sensitivity to the formation time.
\end{enumerate}

\section{Reconstructing $g \to c\bar{c}$ splittings from final state hadrons}\label{sec:reconstruction}

In this Section we will demonstrate the fidelity with which we can access information about the kinematics of the $g \to c\bar{c}$ splitting from simulated hadronic final states. We start by giving details about the Monte Carlo event generation in Section~\ref{sec:event_generation} and the jet substructure analysis techniques employed for reconstruction in Section~\ref{sec:jet_reco}, before showing corresponding simulation results in Section~\ref{sec:rec_results}.

\subsection{Event generation}
\label{sec:event_generation}

For Monte Carlo studies we generate proton-proton collision events at $\sqrt{s}=5.02$~TeV in the Monash tune of \textsc{Pythia}8 with initial state radiation and multi-parton interactions turned on \cite{Bierlich:2022pfr}.
We reconstruct jets using \textsc{FastJet}3 with the anti-$k_t$ algorithm and $R=0.4$ \cite{Cacciari:2011ma}. We assess the importance of hadronisation effects by performing the identical analysis procedure in events with hadronisation off (parton-level) or on (hadron-level). We require either that each reconstructed jet contains a charm-anticharm pair (for parton-level events) or a $D^0$ and $\bar{D}^0$ meson pair (for hadron-level events). We impose a jet-level selection $|\eta^\text{jet}|<1.9$ and no $p_T$ or $\eta$ cuts on tracks. To overlap with experimental feasibility we also require heavy flavour tracks (charm quarks in the parton-level analysis, or $D$ mesons in the hadron-level analysis) to have $p_T^\text{HF}>2$~GeV. If there is more than one possible pairing of $c$-$\bar{c}$ or $D$-$\bar{D}$ pairs inside of a jet, we choose the pair that has the highest transverse momentum when their momenta are summed. For the hadron-level analysis, we turn off decays of $B$ hadrons that could become $D$ mesons, under the assumption that these $B$ decays can be effectively removed experimentally. 

\subsection{Reconstructing the $g \to c\bar{c}$ splitting}
\label{sec:jet_reco}

Since heavy flavour is preserved in QCD splittings, we can reconstruct $g \to c\bar{c}$ splittings even if they happen deep in the parton shower where the gluon energy can be much less than the jet energy. This provides a new type of information compared to standard jet substructure which identifies for example the first hard splitting. However, it also provides a challenge for jet reclustering algorithms, since much higher energy particles from unrelated splittings may contaminate the reconstruction.

As will be clear when we discuss the differences between Fig.~\ref{fig:allplots} and Fig.~\ref{fig:allplots_z0} in the coming section, this can cause  difficulties in reconstructing the splitting kinematics in the regime where the gluon energy $E_g$ is significantly less than the jet energy. Conceptually, this problem can be ameliorated by demanding that the charm quark (or $D^0$ meson) pair carry a sufficiently sizeable fraction of the jet energy. This, however, has the disadvantage that one loses access to the aforementioned splittings where the gluon carries only a small portion of the jet energy. We hence take the somewhat more sophisticated approach of considering only those jets where the $c-\bar{c}$ (or $D^0-\bar{D}^0$) pair is still present in the jet after it is groomed with the SoftDrop algorithm~\cite{Dasgupta:2013ihk, Larkoski:2014wba} with $z_\text{cut} = 0.2$ and $\beta=0$. Since SoftDrop removes soft radiation this ensures that the gluon splitting is within the perturbative part of the parton shower. We stress that after a jet passes the SoftDrop selection criterion, we do the rest of the analysis on the full (ungroomed) jet, which is necessary to gain access to the full splitting kinematics. Fig.~\ref{fig:allplots_z0} demonstrates the necessity of this procedure by showing the splitting kinematics reconstructed for $g \to c\bar{c}$ splittings without this grooming selection. In particular, for the $p_T^{\rm jet}>200\,$GeV$/c$ set it can be seen that the MC has many low energy splitting gluons that are missed by the jet reconstruction algorithms.

\subsubsection{Adaptation of Cambridge/Aachen for heavy flavour}

In principle, the branching history of a parton shower cannot be inferred exactly from the hadrons in a jet. However, since QCD splittings are approximately ordered in angle, the Cambridge/Aachen (C/A) reclustering algorithm reconstructs an approximate branching history by sequentially clustering hadrons in a jet that are nearby in angle. Though angular ordering is not exact, reclustering anti-$k_t$ jets with the C/A algorithm has been successful in extracting the QCD splitting function for the first hard splitting in jets from measurements~\cite{Dasgupta:2013ihk,Larkoski:2014wba,Larkoski:2017bvj}.

For jets containing heavy flavour, this algorithm can be modified to follow the subjets containing the heavy flavour particles in the reconstruction history rather than following the hardest branch. A unique feature of the $g \to c\bar{c}$ splitting is that one can identify the precise splitting where two charm quarks (or charmed mesons) first are in two separate subjets. The kinematics of the charm, anti-charm, and parent gluon in the splitting can then be computed from the four momenta of these subjets.

\subsubsection{Modified FlavourCone algorithm}

An alternative algorithm specifically designed for jets containing one or more heavy flavour hadrons is the FlavourCone algorithm~\cite{Ilten:2017rbd}. In that algorithm, each heavy flavour hadron is the axis of a heavy flavour jet, and all other hadrons are clustered into the closest jet axis provided that this is within a distance equal to the jet radius parameter $R$.

In the original FlavourCone algorithm as proposed in Ref.~\cite{Ilten:2017rbd}, the kinematics of the flavoured jets depends on the jet radius parameter. The key difference in our approach is that we aim to follow the kinematics of the charms as closely as possible.Under the assumption of angular ordering, after a $g \to c\bar{c}$ splitting, all subsequent emissions of either the $c$ or $\bar{c}$ should be at smaller angles than the angle between the $c$-$\bar{c}$ pair. Therefore we use the FlavourCone algorithm with the modification that the radius parameter is fixed to be half the distance between the heavy flavoured hadrons. 
Though this approach may have higher sensitivity to hadronisation effects, for our purposes we find that it provides much more accurate reconstruction of the gluon splitting kinematics.

Under the assumption of exact angular ordering both C/A and FlavourCone should reproduce subjets with the kinematics of the original charm quarks at parton level. Nevertheless, we will see small differences between the two algorithms. 
These differences reflect the fact that any reconstruction of a branching history from the final state is approximate only. In particular, C/A can in general cluster more hadrons into the subjets, since it is a priori only constrained by the jet radius parameter and not by the distance between the flavoured partons or hadrons.

\begin{figure}
    \centering
\includegraphics[width=0.95\textwidth]{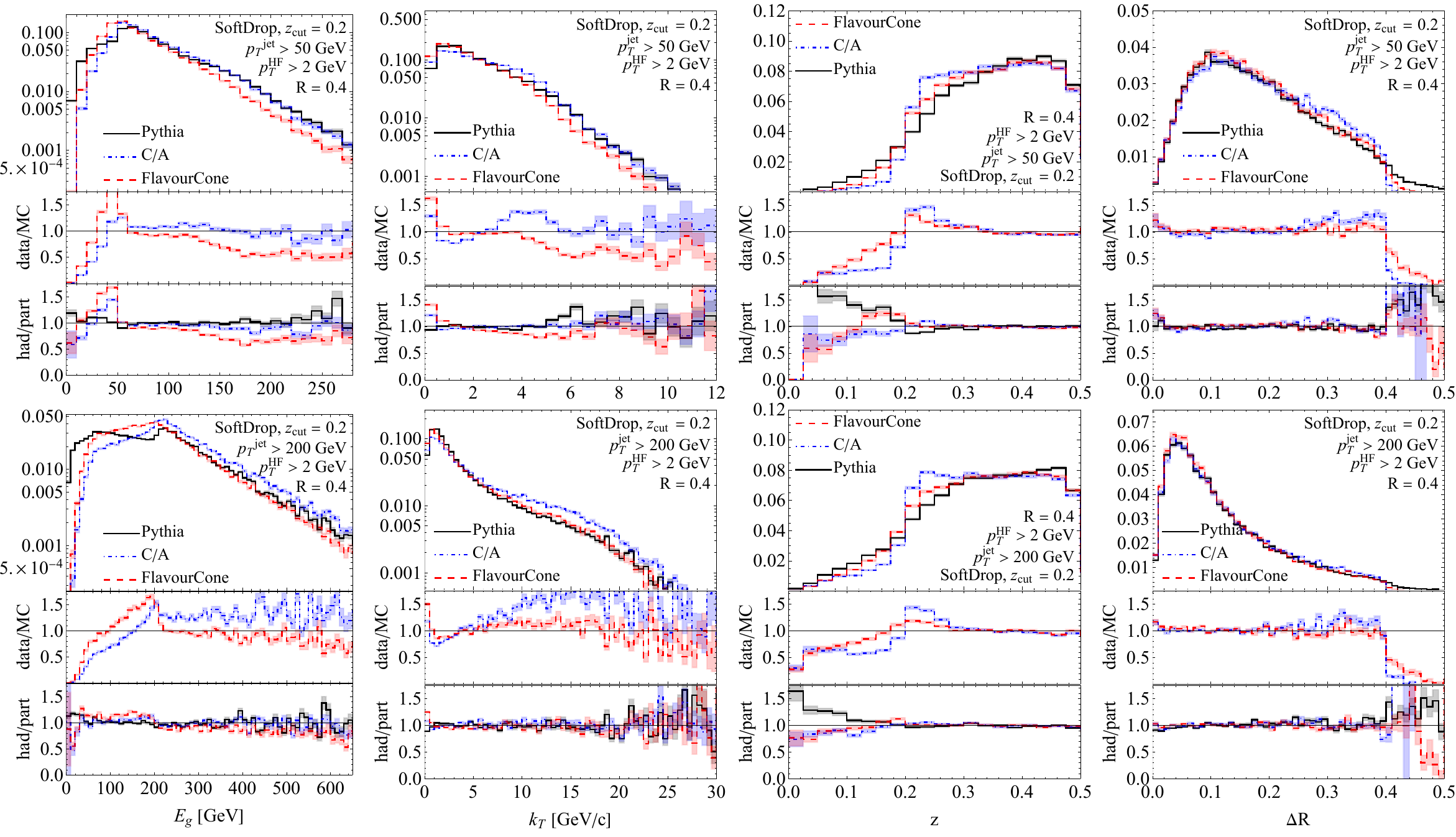}
    \caption{We compare \textsc{Pythia} $g\to c\bar{c}$ splitting kinematics (black) to the reconstructed kinematics from C/A (blue) and FlavourCone (red) for jets with $p_T^\text{jet}>50\,$GeV (upper panel) and $p_T^\text{jet}>200\,$GeV (lower panel). From left to right are the gluon energy $E_g$, charm momentum transverse to the gluon $k_T$, the longitudinal momentum fraction $z$, and the angle between the charms $\Delta R$. Hadronisation effects are quantified by showing the ratio between the hadron and parton level distributions.  }  
    \label{fig:allplots}
\end{figure}

\begin{figure}
    \centering
\includegraphics[width=0.95\textwidth]{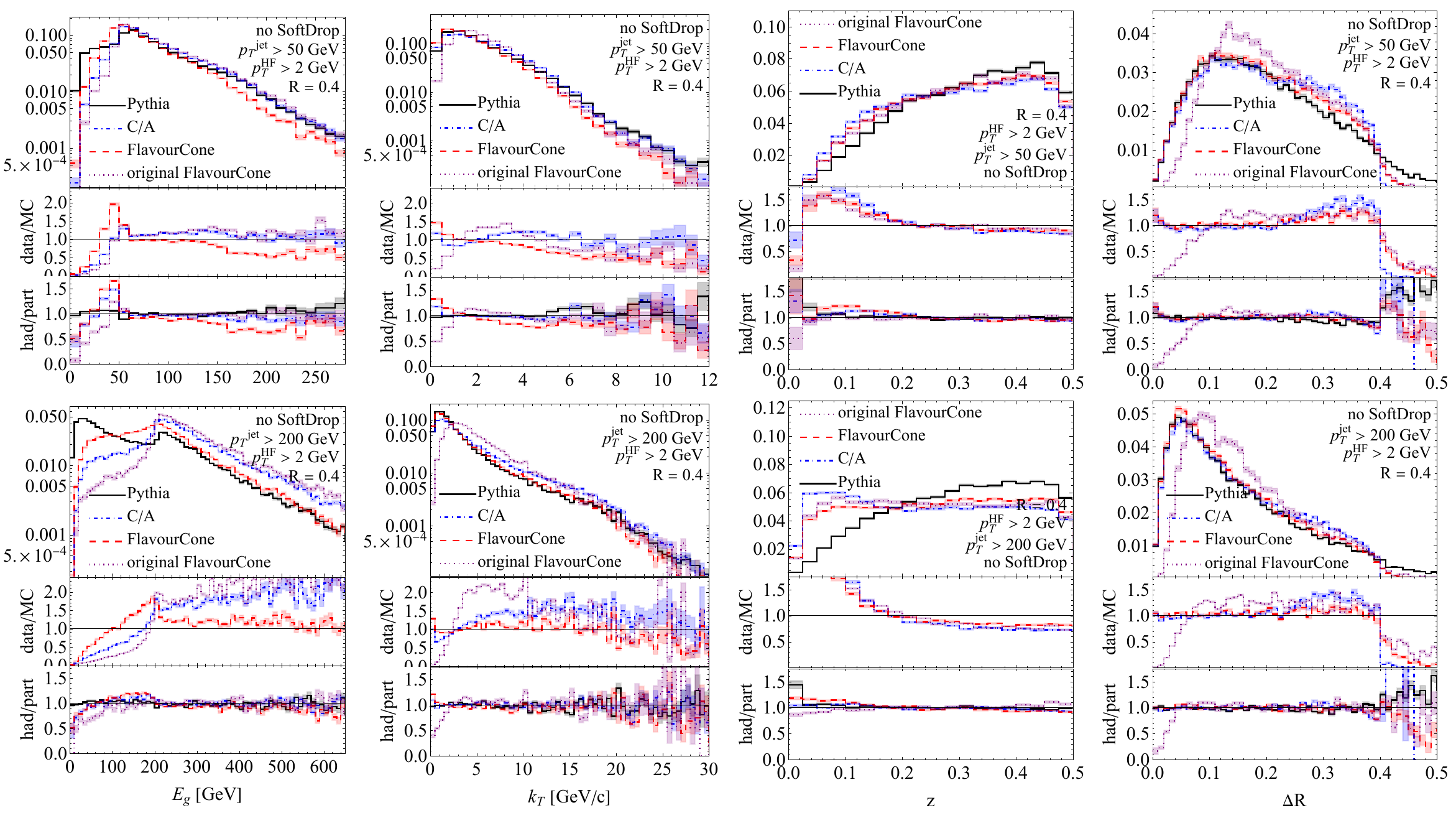}
    \caption{\begin{small}
   To quantify \end{small} the importance of applying our SoftDrop grooming cut and to adapt the FlavourCone jet radius to the distance between the charms we show the same distributions as in Fig.~\ref{fig:allplots}, but now without grooming and also showing the original FlavourCone algorithm. Especially for splittings with a gluon energy much smaller than the jet energy it can be seen that Fig.~\ref{fig:allplots} has a much better agreement with the Monte Carlo. Note that the vertical axes in ratio plots do not necessarily coincide with those in Fig.~\ref{fig:allplots}.}
    \label{fig:allplots_z0}
\end{figure}

\subsubsection{Monte Carlo Validation}

To quantify the accuracy of these techniques, we compare results obtained by  reconstructing the kinematics of the $g \to c\bar{c}$ splitting from the final state to the simulated kinematics of the branching history of a Monte Carlo (MC) parton shower. We repeat that the latter is not measurable directly and that the relation between intermediate stages of that branching history and the hadronic final state is only approximate for the reasons discussed above. Nevertheless, such comparison of simulated ``Monte Carlo truth'' to reconstructed truth provides a valuable baseline to assess the performance of reconstruction techniques. The same strategy has been used in vacuum to establish the feasibility of reconstructing  vacuum splitting functions from the hadronic final state~\cite{Larkoski:2017bvj,Ilten:2017rbd}.

In practice, we perform this comparison as follows.
For jets containing a charm-anticharm pair (for parton-level events) or $D^0$-$\bar{D}^0$ pair (for hadron-level events), we use the MC-internal \textsc{Pythia} shower to identify the highest-$p_T$ $g \to c\bar{c}$ splitting for which the gluon is inside the jet radius. We then extract the kinematics of that splitting from the \textsc{Pythia} shower as Monte Carlo truth. Though they typically coincide, we do not enforce that the daughters of this splitting are necessarily ancestors of the charm-anticharm pair we identify in the final state.  

\subsection{Results}
\label{sec:rec_results}

Fig.~\ref{fig:allplots} shows the kinematics of reconstructed $g \to c\bar{c}$ splittings in jets with $p_T^\text{jet} > 50$~GeV (upper panels) and $p_T^\text{jet} > 200$~GeV (lower panels). We show distributions of the reconstructed kinematic variables of the splitting: the parent (gluon) energy $E_g$, the relative transverse momentum $k_T$ of the splitting, the momentum sharing fraction $z$ and the opening angle $\Delta R$ of the splittees. In principle this leaves some redundancy, since the splitting function can be characterised in terms of three of these variables only; in the collinear limit for example, $k_T = E_g \Delta R z (1-z)$. However, particularly for splittings with $p_T^\text{jet}> 50$~GeV, this relation is often not well-satisfied and so we choose to retain both $k_T$ and $\Delta R$. Reconstructed kinematics using the Cambridge/Aachen (blue dotted) and FlavourCone (red dashed) algorithms are to be compared to the baseline expectation directly from the \textsc{Pythia} shower (black solid). For each figure, the upper panel shows the observable distribution at hadron level, the middle panel shows the ratio of reconstructed quantities to the \textsc{Pythia} baseline, and the lowest panel shows the ratio of each observable distribution between hadron-level and parton-level events with the same selection criteria and analysis procedure. Therefore the middle panel provides an estimate of the size of errors in the reconstruction, while the lower panel indicates the size of hadronisation effects. If the two ratio plots show an equal deviation from unity, the interpretation is that the reconstruction works well at the parton level and that the deviation from the Monte Carlo baseline is primarily due to hadronisation effects. 
Overall, we find that hadronisation effects are relatively mild except at low $z$ (unbalanced splittings) and for angles bigger than the jet radius. There are some moderate hadronisation effects in the $E_g$ distribution, particularly for FlavourCone at lower jet energies. 

Although not perfect, there is a good agreement between the reconstructed kinematics (which are experimentally accessible) and the Monte Carlo ``truth'' kinematics which we are trying to get a handle on. The reconstruction of the energy $E_g$ of the parent gluon branching into $c\bar{c}$ is of special importance for the determination of formation time, since $\tau_f$ depends linearly on $E_g$ by Eq.~\eqref{eq:tauform}. The $E_g$ distribution is seen to be relatively well-captured by our hadronic reconstruction procedures in Fig.~\ref{fig:allplots}, in particular if the gluon energy carries a substantial fraction of the jet energy. 
FlavourCone does significantly better for high $p_T$ jets, whereas C/A performs a bit better for large $E_g$ in low $p_T$ jets. 

For longer formation times, Eq.~\eqref{eq6} implies that we are particularly interested in the kinematic regions for which $k_T$ is small and splittings are  democratic ($z\approx 1/2$). A unique advantage of studying $g\to q\bar{q}$ is that democratic splittings with $z \sim (1-z)$ are the most likely, as opposed to the least likely such as for $g\to gg$ and $q \to q\, g$. Similarly to the case of the $E_g$ distribution, we find also for $k_T$ that FlavourCone does significantly better for high $p_T$ jets, whereas C/A performs a bit better for large $k_T$ in low $p_T$ jets. Also for the $z$ distribution, FlavourCone performs somewhat better. The distribution in the longitudinal momentum fraction of the charm shows a characteristic increase around $z=0.2$ that is stronger for C/A and FlavourCone than for the MC truth. We attribute this to the fact that it is on the final state that we implemented the $z_{\rm cut}=0.2$ SoftDrop criterion, which does not exactly translate to the MC truth. Similarly, the MC truth can have angles $\Delta R$ between the charm pair that are slightly larger than the jet radius $R=0.4$, which are effectively dropped in reconstructed quantities because both charm quarks are required to be in the jet. Nevertheless, the agreement between models and MC truth is still good. 

As mentioned earlier, flavour information allows one to reconstruct splittings deep in the parton shower, but for smaller gluon energies in the range $E_g \ll p_T^\text{jet}$, it becomes increasingly challenging to reconstruct the true parton kinematics. Restricting such a reconstruction to jets in which both charm quarks are still present after grooming with the SoftDrop algorithm is one way of mitigating the risk that partons from unrelated splittings are spuriously included in the kinematic reconstruction of $g\to c\bar{c}$. In the present work, applying this SoftDrop grooming with $z_\text{cut}=0.2$, $\beta=0$ will be our default. It is used in Fig.~\ref{fig:allplots} and it will be used in all subsequent plots, except in Fig.~\ref{fig:allplots_z0}. We include Fig.~\ref{fig:allplots_z0} to illustrate the relevance of the SoftDrop grooming by plotting the same content as in Fig.~\ref{fig:allplots}, except with the difference that we do not reject jets for which the $g \to c\bar{c}$ splitting would be groomed away by doing SoftDrop with $z_\text{cut}=0.2$, $\beta=0$. 

Comparing Fig.~\ref{fig:allplots_z0} with Fig.~\ref{fig:allplots} reveals marked differences that become larger for jet samples of higher energy. Jets with higher energy have typically higher multiplicities and, as a consequence, the likelihood of attributing some of the final state hadrons erroneously is higher. Indeed, especially the gluon energy in high energy jets would often be reconstructed as too large (upper right plot, note also the logarithmic scale). For completeness the figure also includes the original FlavourCone algorithm (using the jet radius $R=0.4$). As expected this leads again to an overestimate of the gluon energy, and also significantly impacts the $k_T$ and $\Delta R$ distributions. Of the proposed algorithms, our modified FlavourCone is the only one for which it is possible to reconstruct the gluon energy with similar accuracy with or without grooming away $g \to c\bar{c}$ splittings that happen on a soft branch.

\section{The formation time in vacuum and medium}
\label{secmed}

\subsection{Formation time in vacuum}
\label{sec3A}

We have illustrated in the previous Section how the kinematics of $g \to c\bar{c}$ splittings inside of jets can be reconstructed from hadron-level events using jet substructure techniques. We next turn to reconstructing the virtualities and formation times of these splittings.

Though hadronisation effects are under relatively good control for reconstructing the splitting in general, as shown in the lower panels of Figs.~\ref{fig:allplots} and ~\ref{fig:allplots_z0}, to access long formation times we seek splittings with low virtuality, which is a region where hadronisation effects can be significant. This can be seen directly from the fact that the virtuality of the gluon cannot be less than the threshold for pair production. At parton-level, this threshold is for the production of two charm quarks ($2m_c = 2.56\,$GeV), while at hadron level it is for the production of the two lightest charmed mesons ($2m_D = 3.73\,$GeV). This is certainly not the only effect of hadronization, but illustrates that hadronization effects can be large close to threshold.

To facilitate comparisons in this Section, we remove this ``trivial'' hadronization effect by shifting parton-level virtuality distributions by the difference between $2 m_D$ and $2 m_c$. In practice, the charm quark mass used in \textsc{Pythia} is $1.5\,$GeV, so we shift parton-level virtuality distributions by $0.73\,$GeV. This is indicated in the plot labels with ``\textsc{Pythia} $(+0.73)$'' and impacts both the virtuality distributions and (indirectly) the formation time distributions for parton-level events.
\begin{figure}
    \centering
    \includegraphics[width=\textwidth]{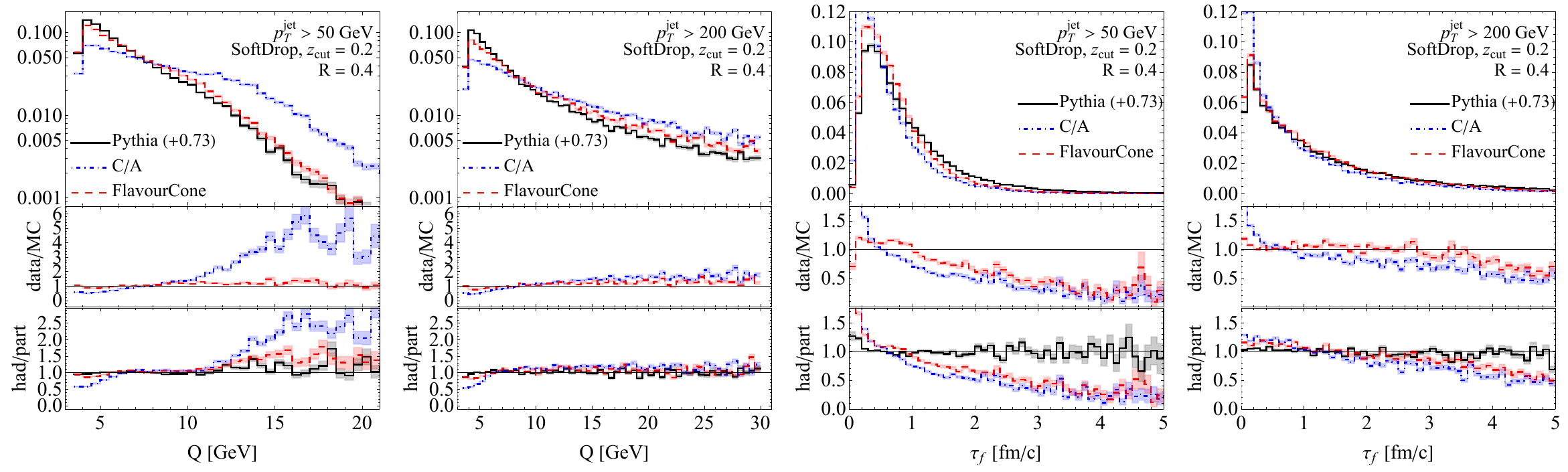}
    \caption{We show the virtuality (left) and formation time (defined in Eq.~\eqref{eq:tauform},  right) for the same samples as in Fig.~\ref{fig:allplots}. In higher $p_T$ jets, the distribution of $c\bar{c}$ pairs tends to be more boosted and of higher virtuality. The interplay of both effects is seen to lead to a broader $\tau_f$ distribution. 
      Hadron level FlavourCone again provides a good estimate of MC truth, provided the virtuality is shifted by the difference in charm and meson mass ($0.73\,$GeV). Reclustering methods provide an accurate reconstruction of the formation time distribution in parton-level events, but hadronisation effects are significant.}
    \label{fig:formationtime}
\end{figure}

In Fig.~\ref{fig:formationtime}, we show the normalised virtuality (left) and formation time distributions (right) of the reconstructed $g\to c\bar{c}$ splittings for the samples from Fig.~\ref{fig:allplots}. In practice, the virtuality is defined as the jet mass of the subjet associated with the splitting (for blue and red curves), or the invariant mass of the pair of daughters as defined from \textsc{Pythia} (black curves). For the distribution in virtuality, a marked difference between FlavourCone and C/A is apparent. At first glance this may be surprising, given the relation between $k_T$, $z$, and the virtuality $Q$, and the relatively good agreement between both C/A and FlavourCone for the $k_T$ distributions in Fig.~\ref{fig:allplots}. The reason is that C/A, as compared to FlavourCone, typically clusters more hadrons in its subjets, which leads to an overestimate of the virtuality. 

For each individual $g\to c\bar{c}$ splitting, the formation time $\tau_f$ results from the combination of the virtuality $Q^2$ plotted in the upper panel of Fig.~\ref{fig:formationtime}, and the gluon energy $E_g$. The latter determines the boost factor $E_g/Q$ by which the average decay time in the parent parton rest frame is boosted. For the samples presented in Fig.~\ref{fig:allplots} the gluon energy can be increased by focusing on higher energy jets. At the same time, however, those gluons also have a wider virtuality distribution. These two effects can partially cancel each other in the formation time. We are mostly interested in how well we can reconstruct the formation time distribution from our experimentally accessible jet reconstruction algorithms. As seen in Fig.~\ref{fig:formationtime}, this reconstruction is under better control if performed in jet samples of higher jet energy $p_T^{\rm jet} > 200$ GeV. We note that the FlavourCone algorithm performs within 25\% for the $p_T^{\rm jet} > 50$ GeV sample for the dominant part of the distribution that has $\tau_f < 2\,$fm$/c$.
The correspondence between the data/MC and hadron/parton curves indicates that the bulk of the uncertainty in the formation time distributions is not due to reconstruction inefficiencies but due to hadronisation effects. In fact, the formation time distribution reconstructed from parton-level events very closely matches the MC baseline. 
\begin{figure}
    \centering
    \includegraphics[width=0.8\textwidth]{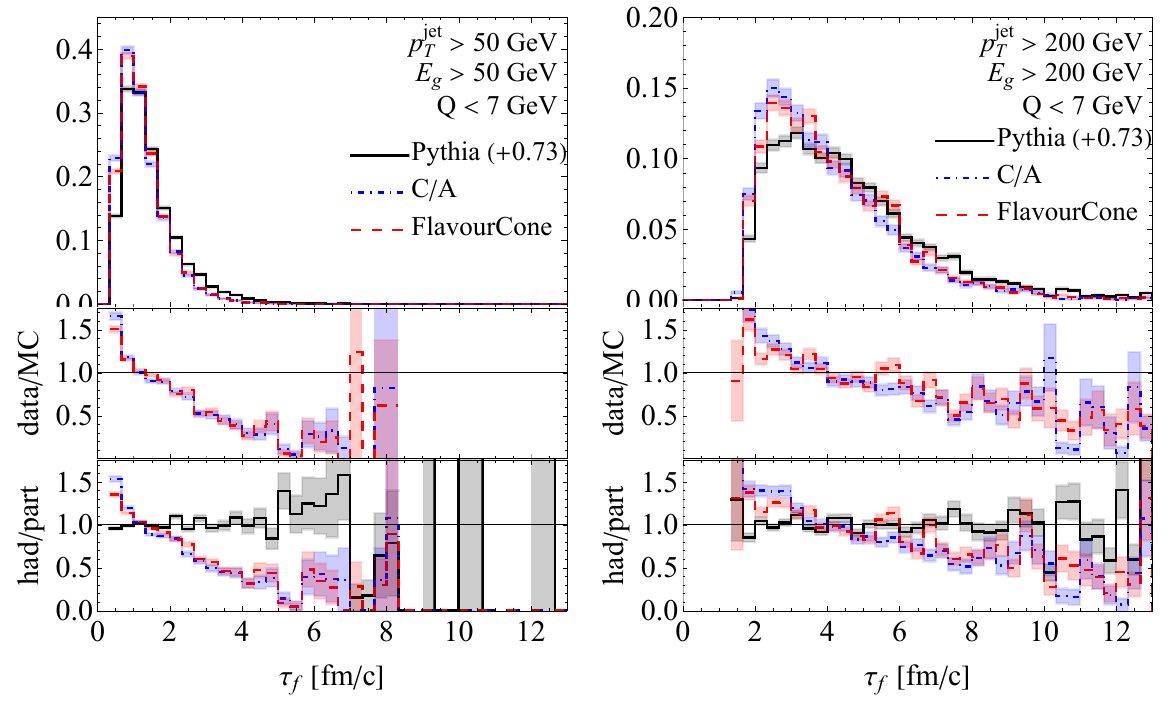}
    \caption{Since our modified jet clustering gives access to the gluon kinematics it is possible to put cuts on hadron level $Q$ and $E_g$ to select splittings with formation times much longer than in Fig.~\ref{fig:formationtime}. In particular, splittings where the gluon is high-energy but with virtuality 
    $Q<7\, {\rm GeV}$ split only late in the medium evolution, with formation times $\tau_f > 2$~fm (right panel). 
    The distributions shown here are self-normalized, since the virtuality and gluon energy selections lead to somewhat different samples for the different reconstruction techniques.}
    \label{fig:formationtimeQcut}
\end{figure}

As discussed at the end of our introductory chapter, to design an experimental test of formation time, it is not sufficient to demonstrate that one can reconstruct $\tau_f$. If one would reconstruct only formation times that are much smaller than the in-medium path length, $\tau_f \ll L$, then the interference factor \eqref{eq2} would be essentially the same for all splittings, and one would observe medium modifications but would not be sensitive to their formation time dependence. It is therefore important to get access to samples of sufficiently long formation times. This is an additional challenge since the $\tau_f$-distributions shown in Fig.~\ref{fig:formationtime} peak at formation times well below 1 ${\rm fm/c}$. The question arises how one can select in a controlled way a sample of sufficiently large formation times that is drawn from the tails of the distributions shown in Fig.~\ref{fig:formationtime}. 

In Fig.~\ref{fig:formationtimeQcut}, we show how a sample of splittings with large formation time can be obtained by restricting the $c\bar{c}$ pair to be produced almost on-shell. We require splittings to have virtuality $Q<7\, {\rm GeV}$ with gluon energy greater than or equal to our jet $p_T$ cut. This selection has the very nice property for our purposes that there is a threshold in the formation time, which can be varied over a substantial fraction of the lifetime of the medium with kinematics that could reasonably be accessible at the LHC. In particular, the right panel shows a selection of splittings that are formed no earlier than $2 \text{ fm}/c$, which is approximately half of the full in-medium path length.
Importantly, even though this is a rather specific kinematic selection, both the C/A and FlavourCone algorithms still describe the MC truth relatively well. We also note that differences between MC and reconstructed formation time distributions in Fig.~\ref{fig:formationtimeQcut} arise almost entirely from hadronisation effects, not from issues with the reconstruction procedures themselves.

\subsection{Formation time in medium}
\label{sec3b}
So far, we have discussed the fidelity with which the formation time \eqref{eq:tauform} of $g\to c\bar{c}$ splittings can be reconstructed from hadronic final states. We have established that such reconstruction can give access to formation times that are sufficiently long to be commensurate with the typical in-medium path lengths in ultra-relativistic heavy ion collisions. We now turn to studying the third critical issue mentioned at the end of the introductory section. Namely, we ask whether the 
formation time dependence of medium modification can be sufficiently sizeable to leave measurable traces.

To address this question, we simulate samples of medium-modified $g\to c\bar{c}$ splittings by reweighting samples of vacuum splittings according to the prescription of Refs.~\cite{Attems:2022otp,Attems:2022ubu}. 
For the medium modification of $g\to c\bar{c}$, we use the path integral formula (2.2) of Ref.~\cite{Attems:2022ubu} and we evaluate it in the multiple soft scattering (harmonic oscillator) approximation. As discussed in Ref.~\cite{Attems:2022ubu}, the results obtained in this harmonic oscillator approximation of $g \to c\bar{c}$ show strong commonalities with the results of the $N=1$ opacity expansion~\eqref{eq4}, they show similar medium modifications in numerical evaluations, and the conclusions of the present work will therefore not depend on the chosen approximation scheme. 
Medium properties enter the multiple soft scattering approximation via the jet quenching parameter $\hat{q}$ which characterises the average squared momentum transferred per unit path length between partonic projectile and medium. This jet quenching parameter has been constrained from data in several model comparisons~\cite{Huss:2020whe,JET:2013cls}. In the present exploratory study, we use the value $\hat{q}\, L = 5\, {\rm GeV}^2$ that lies in the phenomenologically-favoured range. We will show in the coming section that this value results in good agreement between our energy loss model and the measured ATLAS inclusive jet $R_{AA}$.
\begin{figure}
    \centering
    \includegraphics[width=0.8\textwidth]{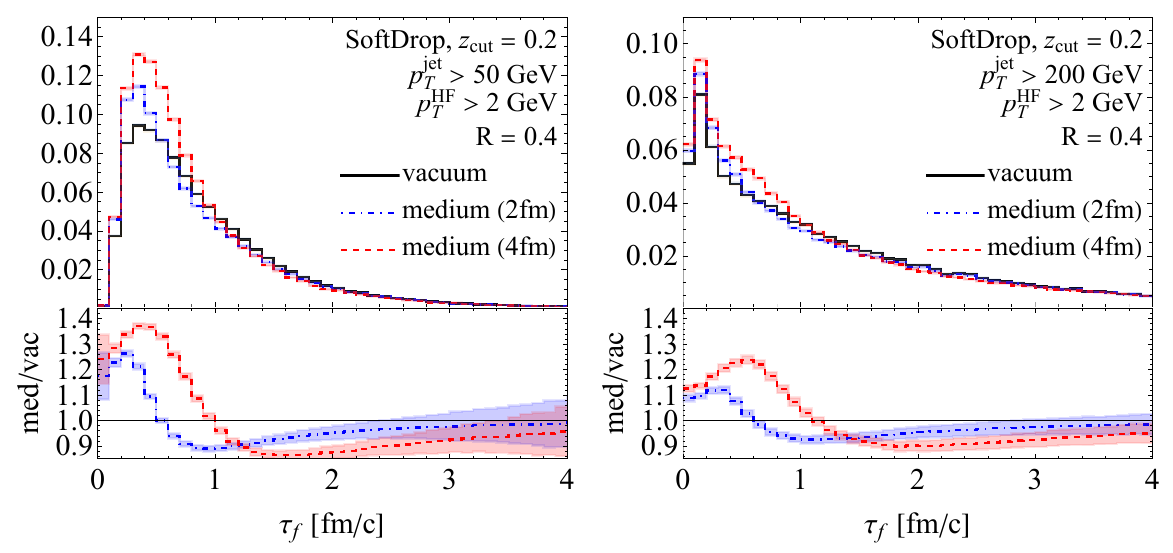}
    \caption{ 
    Using the reweighting prescription of \cite{Attems:2022otp} we show the medium modified $\tau_f$ distributions for a medium length of $L = 2$~fm (dotted) and $4\,$fm (dashed), while keeping $\hat{q}\, L = 5\, {\rm GeV}^2$ fixed. Both vacuum and medium distributions are normalised to the vacuum distribution. The medium $\tau_f$ distributions show an enhancement, momentum broadening leading to smaller $\tau_f$ (see \eqref{eq:tauform}), and are characteristically unmodified for $\tau_f \gtrsim L$.}
    \label{fig:formtimemediummod}
\end{figure}

The medium-modified $g\to c\bar{c}$ in the soft multiple scattering approximation can be written in terms of one dimensionful quantity $\hat{q}\, L$, the dimensionless ratio 
\begin{equation}\label{eq:eg}
    e_g = \frac{2E_g}{\hat{q}L}\frac{1}{L}\,,
\end{equation}
and dimensionless quantities that characterise the transverse momentum and longitudinal momentum fraction (see also section 4.1 of Ref.~\cite{Attems:2022ubu}).
Here $e_g$ can be regarded as the ratio of a formation time $\tfrac{2E_g}{\hat{q}L}$ and $L$. Indeed, $e_g$ plays a role similar to that of the ratio $\tau_f/L$ in the interference factor \eqref{eq2} of the $N=1$ opacity expansion, namely it interpolates between the totally incoherent ($e_g \ll 1$, or $L\gg \tau_f$ at $\hat{q}L$ fixed) and the totally coherent ($e_g \gg 1$, or $L\ll \tau_f$) limiting cases. Of course, \eqref{eq:tauform} and \eqref{eq:eg} are not exactly equivalent since \eqref{eq:tauform} depends on the vacuum kinematics before interaction, whereas $\hat{q}L$ can be thought of as the virtuality gained by the splitting through the medium. Also the small $L$ dependence of the medium modification is different for the harmonic oscillator and the $N=1$ approximation \cite{Caron-Huot:2010qjx}. Away from the totally incoherent limit, $e_g$ in the multiple soft scattering approximation is the only parameter that depends on $L$ while keeping $\hat{q}\, L$ fixed. This is in close analogy to the case of the $N=1$ opacity expansion where, once $n_0\, \sigma_{\rm el}\, L$ is fixed, the 
only $L$ dependence is in the interference factor $\tau_f / L$. This commonality suggests to look for formation time effects by keeping $\hat{q}\, L$ fixed and varying $L$. 

Fig.~\ref{fig:formtimemediummod} shows the result of a study that follows this strategy. We start from  the $\tau_f$ distributions of the two samples of $g\to c\bar{c}$ vacuum splitting functions considered previously. For each splitting in these samples, we calculate the medium modification for two different QGP scenarios that transfer both the same squared transverse momentum $\hat{q}L = 5\, {\rm GeV}^2$ between partonic splitting and QGP, and therefore that share the same incoherent limit. The two scenarios differ by transferring this transverse momentum either within an in-medium path length $L=2$ fm, or within a path length $L=4$ fm that is twice as large. As Fig.~\ref{fig:formtimemediummod} shows explicitly, the medium modification of the kinematic distribution of $g\to c\bar{c}$ splittings shows a characteristic $\tau_f$-dependence that can be varied by changing $L$ with fixed $\hat{q}L$. In particular, if $L$ is increased by a factor 2, characteristic features of the medium modification extend to values of $\tau_f$ that are approximately twice as large. This is a tell-tale sign of a spatial interpretation of $\tau_f$. 

Fig.~\ref{fig:formtimemediummod} is the main conceptual result of the present work. The $\tau_f$ distributions in this figure are constructed based on hadron-level information, and their medium modification clearly imprints the formation time of the splitting. However, the environment is as clean as possible, with medium modification of only the $g \to c\bar{c}$ splitting and no modification of the rest of the jet. The rest of this manuscript is dedicated to demonstrating that formation time effects remain measurable under more realistic experimental scenarios.

We caution that the detailed shape of the $\tau_f$-distributions in Fig.~\ref{fig:formtimemediummod} is complicated to understand on physical grounds as it depends on several competing effects. Fig.~\ref{fig:formtimemediummod} shows a medium modification plotted against the kinematic quantity $\tau_f = \tfrac{2E_g}{Q^2}$ reconstructed from the final state. This is subtly different from Eq.~\eqref{eq:tauform}, where $Q$ is the virtuality before interactions with the medium.
Medium-induced transverse momentum broadening tends to increase the invariant mass $Q^2$ of the $c\bar{c}$-pairs 
in the final state and this medium effect therefore shifts the probability distribution to smaller values of $\tau_f$. This is consistent with the medium-induced enhancement at small $\tau_f$, seen in Fig.~\ref{fig:formtimemediummod}. However, the formalism allows also for additional medium-induced splittings of gluons that would not split in the vacuum~\cite{Attems:2022otp}. On physical grounds, one may expect that these additional splittings are predominantly of small invariant mass and thus of sizeable formation time which could be a competing effect. As we do not have a simple physical picture or estimate of how these different physical effects compete, we can only say that this competition is made precise in the formulas derived in \cite{Attems:2022otp} and that the curves shown in Fig.~\ref{fig:formtimemediummod} arise from evaluating those formulas. We therefore restrict our physical argumentation to the simple and robust observation that for fixed $\hat{q}L$, the medium modification of the $\tau_f$ distribution shows a characteristic dependence on in-medium path length $L$ that is indicative of a spatial interpretation of $\tau_f$.

\subsection{Effects of medium-induced jet energy loss}
\label{subeloss}

\begin{figure}
    \centering
\includegraphics[width=0.5\textwidth]{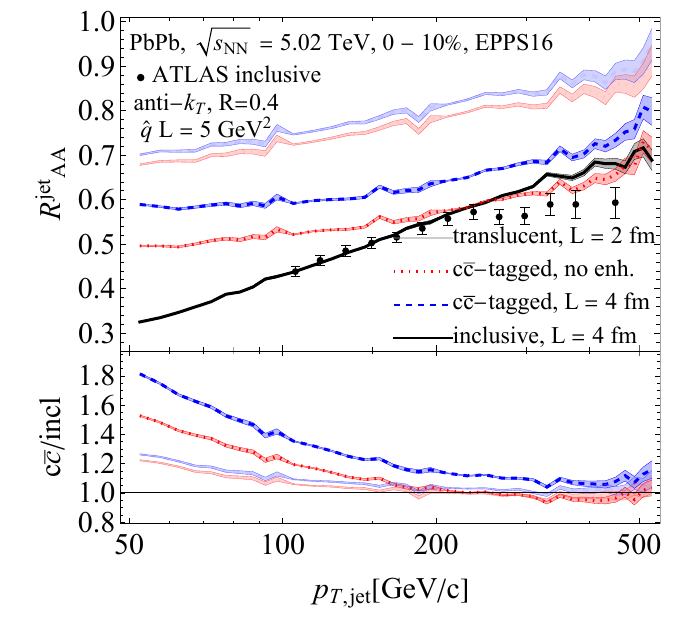}
\includegraphics[width=0.48\textwidth]{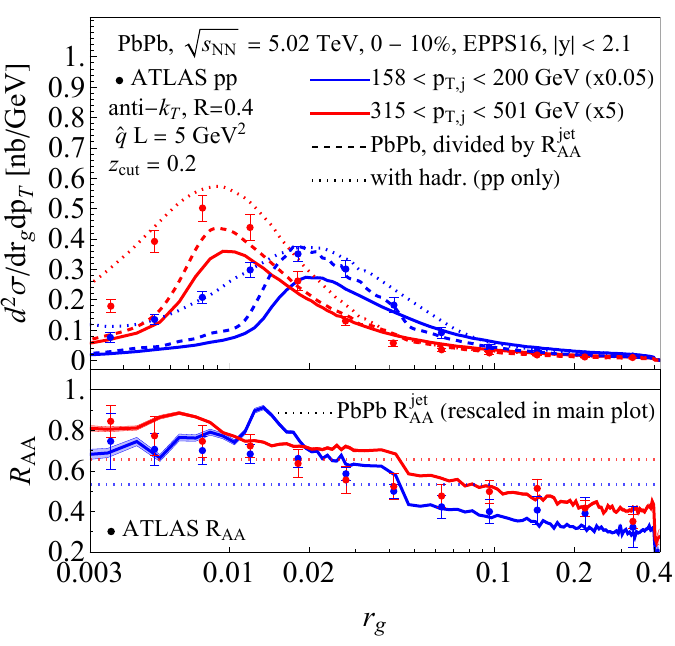}
    \caption{The left panel shows the nuclear modification factor $R_{AA}$ for inclusive jets (black curve) and for $c\bar{c}$-tagged jets with and without medium-modified $g \to c\bar{c}$ splitting function (blue and red curves, respectively). All results are for the jet energy loss model described in the text with $\hat{q}L = 5\, {\rm GeV}^2/{\rm fm}$ fixed and $L=4$ (opaque curves) and $L=2$ (partially translucent). The $R_{AA}$ calculated for inclusive jets shows reasonable agreement with the measurement by ATLAS \cite{ATLAS:2018gwx}. The right panel shows results with the same jet energy loss prescription but for
    the $r_g$ distribution of SoftDrop jets with $\beta=0$ and $z_{\rm cut}=0.2$, compared to ATLAS data \cite{ATLAS:2022vii} for both vacuum ($pp$) and medium modified jets (PbPb at $0-10\%$ centrality). Only ATLAS $pp$ data is shown in the upper panel, while the ratio of PbPb to $pp$ is shown in the lower panel. For improved readability, medium modified jets in the upper panel are rescaled by the $R_{\rm AA}$ (dotted lines in ratio plot) so that their integral matches the vacuum result. Wider jets lose more energy in a medium than narrow jets; in this simple energy loss model this sets in rather sharply at $\theta_c = 2 / \sqrt{\hat{q} L^3} = 0.045 $~\cite{Barata:2023bhh}. } 
    \label{fig:RAA}
\end{figure}

In practice, measuring the medium modification of the $\tau_f$ distribution in  
Fig.~\ref{fig:formtimemediummod} requires reconstructing the kinematics of $g\to c\bar{c}$ splittings in the high-multiplicity environment of ultra-relativistic heavy ion collisions. Measurements of the medium modification of partonic splitting functions have been made in heavy-ion collisions~\cite{CMS:2017qlm, ALargeIonColliderExperiment:2021mqf}, but getting experimental access to the subtle $\tau_f$-dependence of that medium modification has not been attempted yet.
A major question is how jet energy loss may affect the reconstruction of the $\tau_f$ distribution of samples of $g\to c\bar{c}$ splittings. The reweighting procedure used to generate medium effects in Fig.~\ref{fig:formtimemediummod} does not address this issue since it reweights only the $g\to c\bar{c}$ splitting function and it does not account for the dominant medium modifications of the parton shower that arise from the medium modification of $q\to q\, g$ and $g\to gg$ splittings. To further corroborate our conclusions, we therefore include here a first exploratory study of the impact of jet energy loss on our results.

\subsubsection{Jet energy loss prescription and validation}
We implement a simple jet energy loss prescription similar to and inspired by Eq. (3.18) in \cite{Barata:2023bhh}. 
This prescription is based on the picture that energy loss of single partons is separated into two physical regimes: at high energies, energy loss is dominated by semi-hard perturbative gluon emissions that go outside of the jet cone, while at low energies the in-medium emission probability is large and partons undergo turbulent thermalization. Heuristically, this equates to evaluating the quenching weights $Q_i$ in these two regimes as~\cite{Barata:2023bhh}
\begin{eqnarray}
Q_i &=& \exp \Bigl\{ -\int_T^{\omega_s} d\omega \int d^2 {\bf k} \left(\frac{d \mathcal{P}_i^\text{med}}{d\omega d^2{\bf k}}\right)^\text{mini-jet} \left( 1-e^{-\frac{n\omega}{p_T}} \right) 
\nonumber \\
&& \qquad \qquad  -\int_{\omega_s}^{\infty} d\omega \int d^2 {\bf k} \left(\frac{d \mathcal{P}_i^\text{med}}{d\omega d^2{\bf k}}\right)^\text{pert} \left( 1-e^{-\frac{n\omega}{p_T}} \right) \Bigl\}\,,
\label{eq:eloss}
\end{eqnarray}
where $\omega_s = \omega_c \left( \frac{\alpha_s^\text{med} N_c}{\pi} \right)^2$ is the scale where the in-medium gluon emission probability becomes order one, which defines the transition between regimes where the energy loss is dominated by turbulent thermalization or perturbative gluon emission. Explicit formulas for the in-medium rates can be found in \cite{Barata:2023bhh}.

The original prescription of \cite{Barata:2023bhh} was developed to calculate the energy loss of a two parton system, with the partons losing energy independently only if they are separated by an angle larger than $\theta_c = 2 / \sqrt{\hat{q} L^3}$. We extend this prescription slightly to apply it to the energy loss of a jet with an arbitrary number of partons. In a parton level event, all particles in a jet are grouped into clusters of size $\theta_c = 2 / \sqrt{\hat{q} L^3}$. Each cluster is assigned a net flavour based on its parton content: clusters with only gluons and $q\bar{q}$ pairs are ``gluon clusters'', and any other clusters are ``quark clusters''. The energy loss of each cluster is then determined according to Eq.~\ref{eq:eloss}. In practice, we use the quenching weights by rescaling the momenta of each cluster by a factor $Q_i^{1/n}$, where $n=6$ is the power law for the vacuum jet spectrum. The 3-momentum of each particle is rescaled by the energy loss weight of its cluster, the energy is recomputed, and the resulting jet is the medium-modified jet. 

The energy loss prescription discussed above lacks the modelling of phenomenologically-important features such as the embedding of the jets in events of realistic geometry and realistic dynamical evolution. These effects are not thought to be important for quantities like $R_{AA}$, but may have some impacts in more differential observables. Unless otherwise stated, throughout the rest of this work we will take standard values $\hat{q} L = 5 \text{ GeV}^2$ with medium lengths $L=4$~fm (for phenomenological comparisons) and $L=2$~fm. We use the EPPS16 parametrisation of the nuclear parton distribution functions \cite{Eskola:2016oht}. For simplicity in assigning the net cluster flavour, results involving energy loss are shown for parton-level events.
Despite its simplicity, we find that this prescription accounts reasonably well for the $p_T$-dependence and overall suppression of inclusive $R=0.4$ jets measured by the ATLAS collaboration in $\sqrt{s_{\rm NN}}=5.02$~TeV 0-10\% central PbPb collisions at the LHC~\cite{ATLAS:2022vii}, shown by the black curves in the left panel of Fig.~\ref{fig:RAA}. 

The left panel of Fig.~\ref{fig:RAA} also shows the medium modification of $c\bar{c}-$tagged jets using the same jet energy loss prescription, and either including (blue) or not including (red) the medium modification of the $g \to c\bar{c}$ splitting. For comparison to the inclusive case, in this figure we do not require any $p_T$ cut on the $c\bar{c}$ pair, or require it to pass the SoftDrop condition, since these requirements can significantly impact the yield of $c\bar{c}$-tagged jets. At very high $p_T$, where inclusive jets are predominantly quark jets, the difference between the $R_{\rm AA}$ for inclusive jets and $c\bar{c}-$tagged jets results mainly from the enhancement of the $g \to c\bar{c}$ splitting.  At low $p_T$, inclusive jets are predominantly gluon jets. In this kinematic region, the $R_{\rm AA}$ is significantly less suppressed for $c\bar{c}-$tagged jets, consistent with the picture that compared to $g\to gg$, the daughters in $g\to c\bar{c}$ splittings lose less energy. For comparison the left panel of Fig.~\ref{fig:RAA} also shows results for $L = 2$~fm (keeping $\hat{q}L = 5\,$GeV$^2$ fixed). The inclusive spectrum for this reduced in-medium path length would no longer be phenomenologically viable. 

To further validate the utility of our energy loss model, in the right panel of Fig.~\ref{fig:RAA} we compare the results of our model to the ATLAS measurement~\cite{ATLAS:2022vii} of the SoftDrop groomed $r_g$ distribution and its medium modification for inclusive jets. The ATLAS measurement in $pp$ shows reasonable agreement with \textsc{Pythia} at large $r_g$. At small $r_g$ hadronisation effects are significant, as seen by the deviation between \textsc{Pythia} simulations at parton- and hadron-level (solid and dotted curves, respectively). In the lower panels we show the ratios of parton-level events with and without energy loss. Wider jets lose more energy, both in our model and in measurements, with a steep transition in the energy loss occurring at $\theta_c = 0.045$ in our model due to the simplicity of fixing $L$ and $\hat{q}L$ -- more realistic simulations with fluctuating geometry would smoothen this effect. 
  
As this model accounts for the correct magnitude and reasonable angular dependence of jet energy loss, it can provide insight into the role of energy loss in the phenomenology of the $g \to c\bar{c}$ splitting, and in particular into the question of whether the signatures of a spatio-temporal interpretation of formation time found in Fig.~\ref{fig:formtimemediummod} can be expected to remain experimentally accessible in the presence of parton energy loss. 

\subsubsection{Utility of $g \to c\bar{c}$ tagging for measuring medium-induced momentum broadening}

Before returning in the next section to the formation-time dependence of medium effects, we will take a brief detour to highlight a major advantage of $c\bar{c}$-tagged jets for measuring medium-induced momentum broadening. 
A well-known effect in heavy-ion collisions is that medium-induced energy loss narrows the opening angle distribution of jets through selection bias effects. In particular, jets in heavy-ion collisions are strongly biased towards those that lost as little energy as possible, which are more likely narrow jets. This effect is flavour dependent, since gluon jets fragment earlier and at wider angles and therefore have higher multiplicity and larger energy loss compared to quark jets. One can see this effect explicitly in the right panel of Fig.~\ref{fig:RAA}, where the effect of energy loss alone (particularly at large $r_g$) narrows the $r_g$ distribution in heavy-ion collisions. This presents a serious challenge for phenomenology, since the effects of medium-induced momentum broadening would need to be measured on top of this strongly shifted baseline, and it is difficult to constrain the shift of this baseline without substantial model dependence.

\begin{figure}
    \centering
\includegraphics[width=0.44\textwidth]{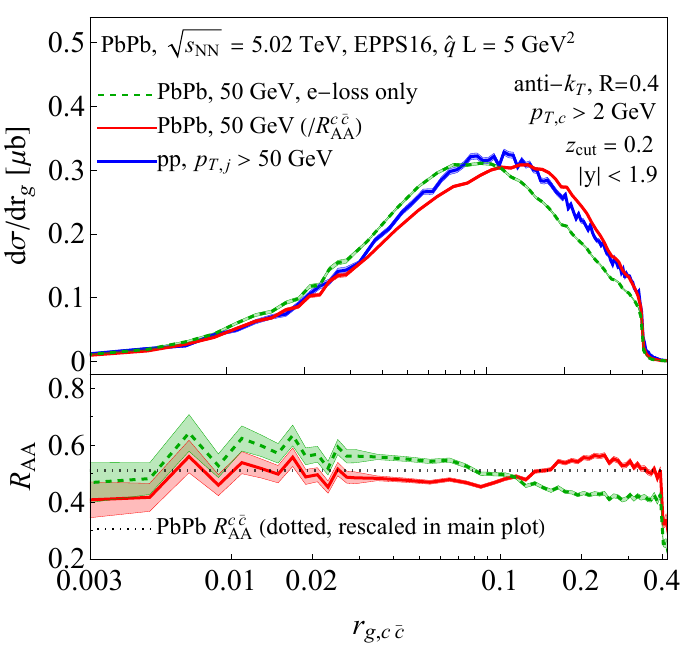}
\includegraphics[width=0.44\textwidth]{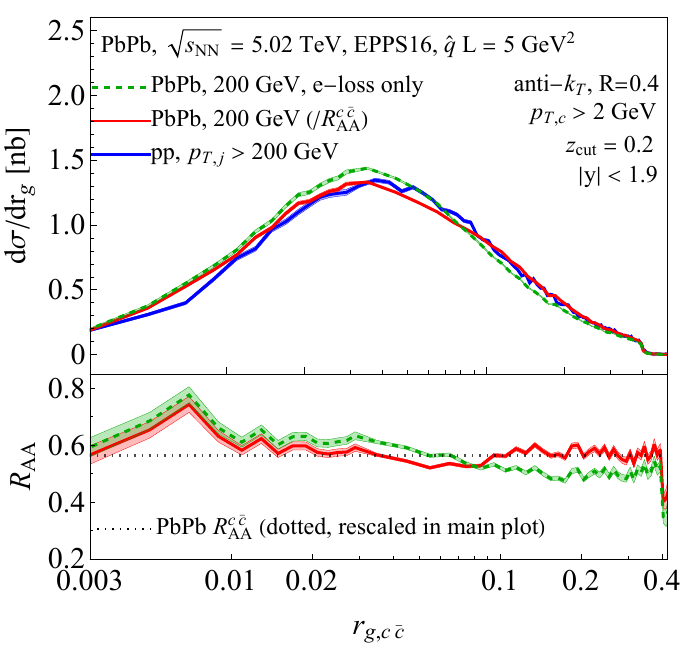}
    \caption{We show the angle between the $c\bar{c}$-pair (as reconstructed using the FlavourCone algorithm) for vacuum ($pp$) and medium modified $c\bar{c}$-tagged jets for jets with $p_T^\text{jet}>50$~GeV (left) and $p_T^\text{jet}>200$~GeV (right). Due to tagging $g \to c\bar{c}$ jets, the effect of energy loss is more moderate compared to that in inclusive jets (only shown in ratio plot), and after including momentum broadening via the medium modification of the $g \to c\bar{c}$ splitting function, we see that a moderate amount of broadening may be observable.
    \label{fig:rgdistr_ccbar}}
\end{figure}

An important consequence of tagging $g \to c\bar{c}$ jets is that it makes it possible to eliminate selection biases arising from the flavour-dependence of jet energy loss by isolating the medium modification of a single type of splitting. This point is illustrated in Fig.~\ref{fig:rgdistr_ccbar}. Compared to the inclusive case in the right panel of Fig.~\ref{fig:RAA}, the variation of the $r_g$ distribution arising from energy loss only is significantly milder for $g \to c\bar{c}$-tagged jets (comparing solid lines in the ratio panel of the right panel of Fig.~\ref{fig:RAA} to green lines in the ratio panels of Fig.~\ref{fig:rgdistr_ccbar}). This dramatic reduction in the selection bias may enable future measurements of medium-induced momentum broadening in $c\bar{c}$-tagged jets, as shown in red in Fig.~\ref{fig:rgdistr_ccbar}.

\subsubsection{Accessing formation time effects}

\begin{figure}
    \centering
\includegraphics[width=0.88\textwidth]{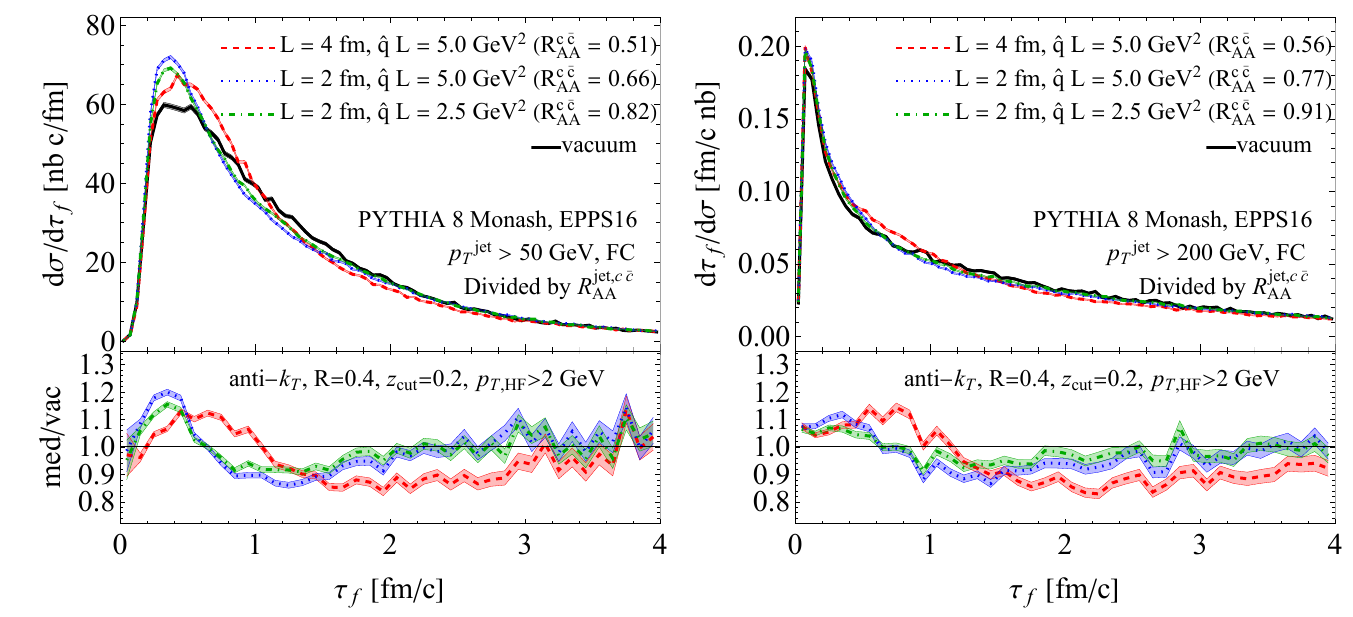}
    \caption{
    We show the formation time distribution as comparable with Fig.~\ref{fig:formtimemediummod} except with energy loss included. The medium modified curves are rescaled by the relevant inclusive jet $R_{\rm AA}$ to account for the reduction in yield due to energy loss. After this rescaling the distributions are remarkably similar. Note that this figure displays a cross section instead of being self-normalised. 
    }
    \label{fig:formtimemedmod_eloss}
\end{figure}

We finally are equipped to address how jet energy loss impacts the observability of the formation-time dependent medium modification shown in Fig.~\ref{fig:formtimemediummod}. In Fig.~\ref{fig:formtimemedmod_eloss}, we show how this formation time distribution is modified if medium-induced jet energy loss is included in the simulation. One immediate effect with respect to the vacuum is a change in the overall yield, as captured by the $R_{\rm AA}$. To make the figures more comparable, the medium modified curves in Fig.~\ref{fig:formtimemedmod_eloss} are rescaled by the relevant $R_{\rm AA}$ of $c\bar{c}$-tagged jets.
After this rescaling, the medium modified formation time distributions shown in red and blue curves are remarkably similar to Fig.~\ref{fig:formtimemediummod}. In particular, if $L$ is increased by a factor of two, the shape of the medium modification ratio stretches again by a factor of about two, as expected for the functional $\tau_f/L$- dependence of a formation time effect. Based on this exploratory study on parton level, we see an experimental opportunity of establishing evidence for the parton formation time by identifying characteristic signatures in the $\tau_f$ distribution of $c\bar{c}$-tagged jets. 

In the red and blue curves of Fig.~\ref{fig:formtimemedmod_eloss}, we have kept $\hat{q}L = 5\, {\rm GeV}^2$ fixed and we have varied $L$. This has a clean theoretical interpretation, since the only $L$ dependence that remains when $\hat{q}L$ is fixed is due to the formation time. In practice, however, while it is reasonably straightforward to change $L$ (for example, via centrality cuts, or in collisions of smaller ions), it may be difficult to keep $\hat{q} L$ fixed, since $\hat{q}$ is an in-medium property that is difficult to vary by a large factor. It is hence worth noting that one may not need to keep $\hat{q}L$ fixed. The green curves in Fig.~\ref{fig:formtimemedmod_eloss} show  the medium modification of the formation time distribution for $L=2$~fm and $L=4$~fm except with $\hat{q}$ fixed, corresponding to $\hat{q}L = 2.5 \text{ GeV}^2$ and $\hat{q}L = 5 \text{ GeV}^2$, respectively. We find the same qualitative formation time effect as seen in the blue curves, namely that the medium modification vanishes above the medium length $L$. This indicates that one does not need to experimentally engineer conditions with $\hat{q}L$ fixed and varying  $L$ to test the spacetime interpretation of $\tau_f$.

\section{Discussion}
There have been numerous experimental and theoretical efforts in recent years to  understand with higher precision the effects of parton energy loss. Most of these developments utilise jets as tools to learn about properties of the QGP with higher precision. As emphasised in the present work, there is also merit in inverting this logic to utilise the QGP as a tool to learn about the parton shower dynamics. In particular, relatively few works so far have considered the possibility of directly testing the assumed space-time picture of a parton shower by studying its embedding in the QGP. The present study is the first that explored this possibility by studying event samples of individual $1\to 2$ splittings reconstructed from the final state in nucleus-nucleus and proton-proton collisions. 
 
We have explored a particular measurable quantity, namely the formation time distribution of samples of reconstructed $g\to c\bar{c}$ splittings. We have established that this quantity satisfies important criteria needed to make the spatio-temporal interpretation of the formation time experimentally testable. In particular, by selecting jets in which a $c\bar{c}$ pair is produced on a sufficiently hard branch, we found that modified C/A and FlavourCone algorithms can access the gluon splitting kinematics with high fidelity. This allows almost direct access to the gluon formation time in vacuum. In the medium, the kinematics of the splitting is modified. We have shown that the resulting formation time-dependence of this medium modification is numerically sizeable, and may be measurable. In addition, there is the experimental and theoretical challenge of disentangling the medium modification of the $g\to c\bar{c}$ splitting from other effects of medium-induced jet energy loss. In the context of a simple jet energy loss prescription, we have shown that characteristic features of the medium-modified shape of the formation time distribution remain visible if jet energy loss is included in the simulation. 
The tell-tale sign of the formation time-dependence is that results depend on the medium path length $L$ when the total momentum transferred $\hat{q}L$ is fixed. Nevertheless, in Fig.~\ref{fig:formtimemedmod_eloss} we note that keeping $\hat{q}$ fixed while varying $L$ can also give access to the effects of the formation time experimentally.

We emphasise the exploratory character of our study.
The formation time-dependence of jet quenching is a rather subtle modulation, while jet quenching itself is a numerically large phenomenon that is observed generically in almost all hadronic high transverse momentum observables. However, the formation time dependence of jet quenching can only be identified if one manages to reconstruct with sufficiently high fidelity those relatively rare branching processes whose formation time is comparable to or larger than the traversed medium path length. Future studies including the modification of the $g \to c\bar{c}$ splitting on equal footing with the modification of other splitting functions, and incorporating the effects of fluctuating expanding geometry in heavy-ion collisions, can put these results on stronger quantitative footing. 

We have demonstrated how an experimentally accessible $\tau_f$-modulation of jet quenching 
manifests itself in the characteristic enhancement and depletion shown in the ratio plots in Figs.~\ref{fig:formtimemediummod} and~\ref{fig:formtimemedmod_eloss}. While undoubtedly experimentally challenging, first measurements
toward testing the QCD formation time with reconstructed parton splittings could provide quantitative access to the formation time dependence of medium interactions, and shed light on the spatial structure of parton branching processes.\\

\noindent
\textbf{Acknowledgments:} 
It is a pleasure to thank Maximilian Attems, Gian Michele Innocenti, Michelangelo Mangano, Aleksas Mazeliauskas, Pier Monni, Gavin Salam, Jesse Thaler and Nima Zardoshti for useful discussions. We are also thankful to Dhanush Hangal and Martin Rybar for providing data points from the ATLAS measurement that are reproduced in Fig.~\ref{fig:RAA}. JB acknowledges support through a University Research Fellowship from the Royal Society under grant URF\textbackslash R1\textbackslash 241231, from the Leverhulme Trust under grant LIP-2020-014, and from the UK Science and Technology Facilities Council (STFC) under grant ST/T000864/1.\\

\noindent
\textbf{Source code:} 
The code to reproduce results in this paper can be found at \url{https://github.com/jasminebrewer/ccbar_substructure}.

\bibliographystyle{JHEP}
\bibliography{ccbar_formation}

\end{document}